\documentclass[%
reprint,
superscriptaddress,
 amsmath,amssymb,
 aps,
]{revtex4-2}

\usepackage{graphicx}
\usepackage{dcolumn}
\usepackage{bm}
\usepackage{bibunits}
\usepackage{hyperref}

\begin{document}

\title{Deep rotating convection generates the polar hexagon on Saturn}

\author{Rakesh Kumar Yadav}
\affiliation{Department of Earth and Planetary Sciences, Harvard University, 02138 Cambridge, USA}
 \email{To whom correspondence should be addressed: rakesh\_yadav@fas.harvard.edu}

\author{Jeremy Bloxham}
\affiliation{Department of Earth and Planetary Sciences, Harvard University, 02138 Cambridge, USA}

\date{\today}

\begin{abstract}
Numerous land and space-based observations have established that Saturn has a persistent hexagonal flow pattern near its north pole. While observations abound, the physics behind its formation is still uncertain. Although several  phenomenological models have been able to reproduce this feature, a self-consistent model for how such a large-scale polygonal jet forms in the highly turbulent atmosphere of Saturn is lacking. Here we present a 3D fully-nonlinear anelastic simulation of deep thermal convection in the outer layers of gas giant planets which spontaneously generates giant polar cyclones, fierce alternating zonal flows, and a high latitude eastward jet with a polygonal pattern. The analysis of the simulation suggests that self-organized turbulence in the form of giant vortices pinches the eastward jet, forming polygonal shapes. We argue that a similar mechanism is responsible for exciting Saturn's hexagonal flow pattern.\\
\\
{\bf Paper published in the {\em Proceedings of the National Academy of Sciences.} \\
\href{https://doi.org/10.1073/pnas.2000317117}{https://doi.org/10.1073/pnas.2000317117} }
\end{abstract}

\maketitle

\section{Introduction}
In 1988, Godfrey \cite{godfrey1988} analyzed the 1981 flyby data from Voyager 2 and reported one of the most visually spectacular features in planetary atmospheres: the presence of a hexagonal pattern in the prograde zonal jet at around 15 degrees away from Saturn's north pole. Since its discovery, Saturn's hexagon (hereafter `hexagon') has been repeatedly observed and we know that at least in the last 40 years or so, the hexagon has been present and relatively unchanged \cite{caldwell1993,sanchez1993}. The hexagon exhibits some dynamical behavior, including drifting slowly in the westward/eastward direction with speeds ranging from -0.06 to 0.01 degrees per day (in Saturn's System III reference frame) \cite{caldwell1993,hueso2019}, although, given the uncertainty associated with Saturn's rotation period \cite{anderson2007,read2009,mankovich2019}, it is rather difficult to infer the actual drift rate of the hexagon. The hexagon also encloses a circumpolar cyclonic (spinning in the planetary rotation direction) vortex that is also known to be a stable feature \cite{sanchez1993,fletcher2008}. We refer the reader to a recent comprehensive review by Sayanagi {\it et al.} \cite{sayanagi2018} for more detail on the observational and modeling history of Saturn's atmosphere.

The existence of such a prominent and stable feature on Saturn gives us an opportunity to test different possibilities for how atmospheric dynamics in Saturn generates such features. Over the years, several models have been proposed. Shortly after the discovery of the hexagon, Allison {\it et al.} \cite{allison1990} argued that the hexagon is essentially a stationary Rossby wave produced by the interaction of the eastward jet with a large anticyclonic vortex to the south of the jet visible in the Voyager 2 data. When Cassini later visited Saturn, this large anticyclonic vortex was no longer present \cite{fletcher2008}, questioning the idea of a forced Rossby wave. On the other hand, S{\'a}nchez-Lavega {\it et al.} \cite{sanchez2014} argue that the hexagon is a stationary unforced Rossby wave that exists on a deep (may be deeper than 10 bars) quasi-geostrophic zonal jet. In the most recent development on the modeling front, Morales-Juber{\'i}as {\it et al.} \cite{morales2015} study how perturbations affect an eastward jet stream. In their model, perturbations evolve into hexagonal-shaped meanders when the jet decays below the 2 bar level. The speed, decay rate, and the curvature of the jet determine the dominant wave number of the meander in this model.

Laboratory experiments have also shed light on the possible mechanisms for hexagon formation. Sommeria {\it et al.} \cite{sommeria1989} performed experiments on a rotating annulus where barotropic zonal jets  were generated using mechanical forcing in the form of mass sources and sinks. Depending on the mass flow rate and the rotation rate of the container, they reported the existence of wavy jets with different azimuthal wavenumbers (from 3 to 8). They interpret these features as Rossby waves excited in the region where the potential vorticity has sharp gradients. Each edge of the wavy perturbation on the jet was accompanied by an adjacent vortex. More recently, Aguiar {\it et al.} \cite{aguiar2010} also reported wavy shapes in a barotropic zonal jet excited in rotating tanks using external forcing. Depending on the rotation rate of the tank and the flow speed of the forced jet, they report wavy features with wavenumbers ranging from 2 to 8. They suggest that these polygonal patterns are the manifestation of a fully developed barotropic instability in a zonal jet. They also observe vortical features adjacent to the polygon edges on the zonal jet.

Considering the various theoretical models, simulations, and laboratory experiments discussed above, the essential idea emerges that jets can become unstable and give rise to polygonal features. However, we note that all these studies either assume a zonal jet or it is generated via external forcing. Furthermore, the deep planetary convection, which might be the fundamental driving force behind the zonal jets, has not been modelled in the earlier studies of hexagon formation. A model of how highly non-linear fluid turbulence self-organises and gives rise to zonal jets with geometrical shapes is  lacking.

The situation is rather different if we take a broader perspective of the Saturn's atmospheric dynamics. The planetary scale alternating zonal jets on Saturn have been studied in great detail and have been generated in a completely spontaneous and self-consistent manner in models incorporating fluid turbulence. Here, two schools of thoughts have developed: on one hand, the alternating jets on gas giant planets are `shallow', existing above 10 bars or so \cite{williams2003,cho1996,liu2010}; on the other hand, zonal jets are `deep', extending to tens of thousands of bars \cite{busse1976,christensen2001,aurnou2001,heimpel2005,kaspi2009}. In this regard, exciting developments were made  recently shortly before Cassini took its final plunge into Saturn (the `Cassini Grand Finale'): the interpretation of the gravity harmonics from the final Cassini orbit hints at Saturnian zonal jets retaining their strength down to at least 100,000 bars \cite{galanti2019}, strongly suggestive of  the deep jets scenario.

The recent Cassini results and the fact that the hexagon has remained stable for the last 40 years or so (unperturbed by the solar radiation forcing through Saturn's year) suggests that it might be a deep rooted feature as previously noted by S{\'a}nchez-Lavega {\it et al.} \cite{sanchez2014}. Following this line of reasoning, here we report a global simulation where our primary aim is to simulate one of the most basic phenomenon happening in the outer layers of Saturn, namely, deep turbulent compressible convection in a rotating spherical shell. Several simulation studies have been conducted in the past to investigate the deep-convection driven atmospheric dynamics of gas and ice giant planets. They have reproduced the equatorial super-rotation \cite{christensen2001,aurnou2001,heimpel2005,kaspi2009}, similar to Saturn and Jupiter, as well as subrotation \cite{aurnou2007,gastine2013}, similar to the ice giants. Several \cite{heimpel2007,gastine2014} have also investigated the properties (number of jets and their strength) of mid to high latitude alternating zonal jets on Saturn and Jupiter. However, none of these studies report Saturn-like polygonal jets.

\begin{figure*}
\centering
\includegraphics[width=0.9\linewidth]{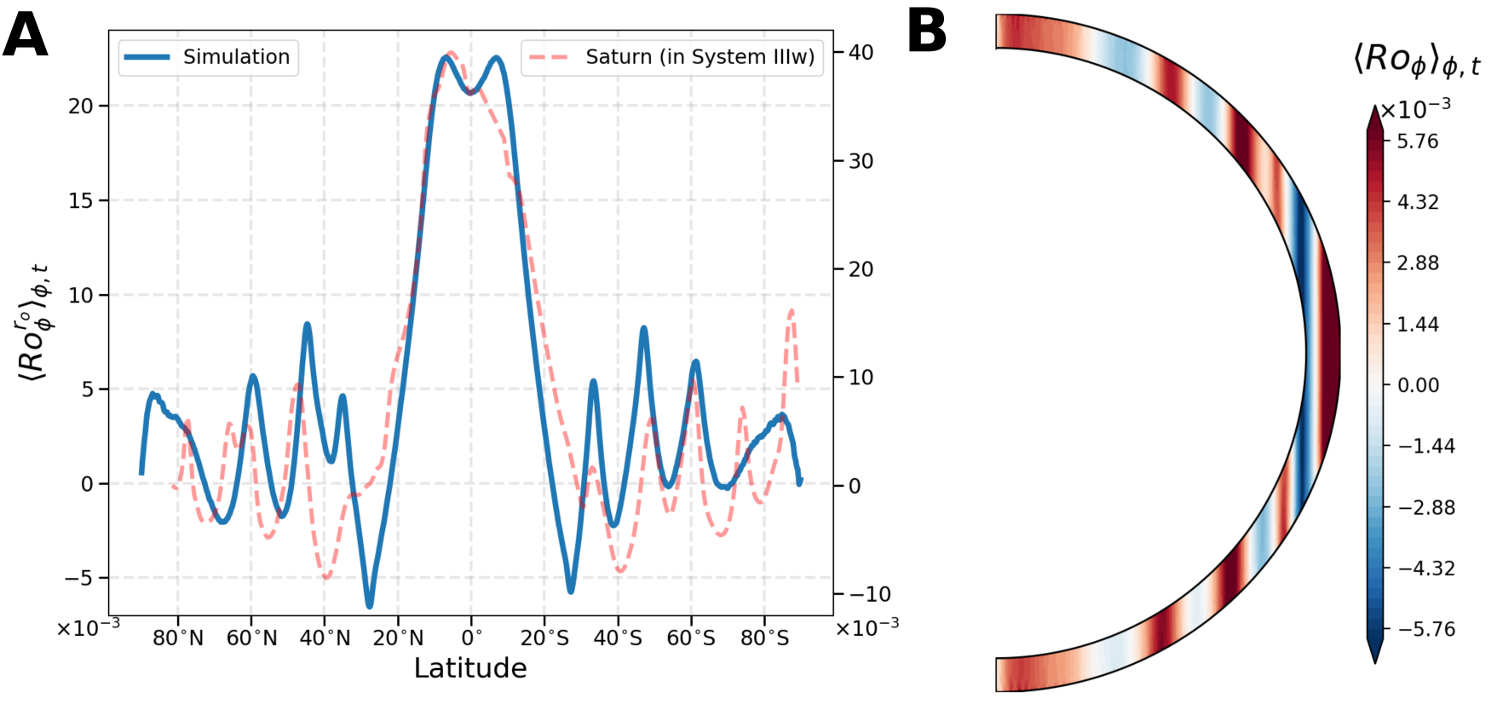}
\caption{Panel {\bf A} shows the azimuthally-averaged zonal flow on the outer boundary of the simulation (left axis) and for Saturn (right axis).  Panel {\bf B} shows the azimuthally-averaged zonal flow on a meridional plane. The simulation data is time-averaged over a few rotations of the shell. The zonal velocity magnitude is given in terms of the Rossby number defined here as $u/(\Omega r_o)$, where $u$ is velocity, $\Omega$ is shell rotation rate, and $r_o$ is outer radius of the shell. The Rossby number for Saturn is calculated using the observed cloud level zonal flow velocities, Saturn's mean radius $5.8232\times10^7$ meters, and the rotation period $1.64\times10^{-4}$ rad/s in System IIIw \cite{read2009}.}
\label{fig1}
\end{figure*}

\begin{figure*}
\centering
\includegraphics[width=0.6\linewidth]{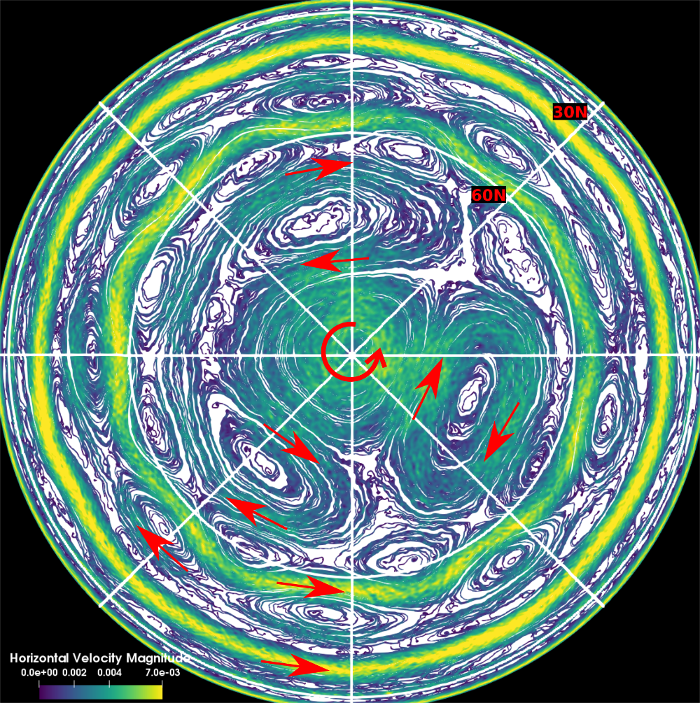}
\caption{Orthographic view of flow stream lines on a spherical surface at radius 0.95$r_o$, as viewed from a north pole vantage point. The stream lines were generated by randomly placing 4000 massless seed particles in the northern hemisphere and then tracing their trajectories governed by the local horizontal flow. The streamlines help us to visualize the {\em instantaneous} velocity structures. Since there are only a finite number of tracer particles, some of the regions on the spherical surface remain untraced. We fill this empty region with white color for clarity. The color of each streamline represents the magnitude of the local horizontal flow velocity in Rossby number. The 30 degree and 60 degree north latitudes are highlighted in the plot using white circles. The red arrows indicated the general flow direction of the stream lines. }
\label{fig2}
\end{figure*}

\section{Method}
We assume that the Saturn's interior consists of a  deep dynamo region, where strong magnetic fields are generated that inhibit  strong zonal flows, and an outer `atmospheric' layer where the electrical conductivity of the fluid is low, allowing strong zonal flows. Here, we simulate hydrodynamic convection {\em only} in the outer layer. We assume a spherical shell with inner radius $r_i$ fixed at 0.9$R_S$ (where $R_S$ is Saturn's radius) and the outer radius $r_o$ at $R_S$. The aspect ratio of the shell is then defined as $\eta=r_i/r_o$ which is 0.9 in this case. The shell rotates with angular velocity $\Omega$. We employ the widely used anelastic approximation \cite{braginsky1995,lantz1999} that allows density stratification in the fluid but filters out sound waves. In this approximation, thermodynamic quantities are decomposed as a static background and a small fluctuation $\tilde{x}(r)+x'(r,\theta,\phi)$. Here we assume a density-stratified hydrostatic and adiabatic reference state defined by 
\begin{gather}
\frac{d\tilde{T}}{dr} = -\frac{g}{c_p}
\end{gather}
where $\tilde{T}$ is reference state temperature, $g$ is gravity, and $c_p$ is specific heat at constant pressure; $c_p$ is assumed constant. We assume an ideal gas fluid, giving a polytropic equation of state where the background density and temperature are related by $\tilde{\rho}=\tilde{T}^m$, where $m$ (assumed to be 2) is the polytropic index. Gravity is inversely proportional to $r^2$ ($r$ being the radius) which assumes most of the planetary mass is below 0.9$R_S$ \cite{jones2009,gastine2012}. We refer the reader to Jones and Kuzanyan \cite{jones2009} for a more detailed discussion about the anelastic equations used in the planetary deep-convection community. 

\subsection{Anelastic Equations}
The non-dimensional evolution equation for velocity is:
\begin{gather}
\frac{\partial{\bf u}}{\partial t}+{\bf u}\cdot\nabla\mathbf{u}+ 2\,\hat{z}\times{\bf u}=-\nabla\frac{p}{\tilde{\rho}}+\frac{Ra\, E^{2}}{P_{r}}\, \frac{r^2_o}{r^2}\, s\,\hat{r} +\frac{E}{\tilde{\rho}}\nabla\cdot S, \label{eq:vel} 
\end{gather}
where $p$ is pressure, $\mathbf{u}$ is velocity, $s$ is entropy, $\hat{z}$ and $\hat{r}$ are rotation and radial unit vectors respectively,  
\begin{gather}
S_{ij}=2\tilde{\rho}\left(e_{ij}-\frac{1}{3}\delta_{ij}\nabla\cdot\mathbf{u}\right) 
\end{gather}
is the traceless rate-of-strain tensor with $\delta_{ij}$ being the identity matrix and 
\begin{gather}
e_{ij}=\frac{1}{2}\left(\frac{\partial u_{i}}{\partial x_{j}}+\frac{\partial u_{j}}{\partial x_{i}}\right).  
\end{gather} 
The entropy is governed by
\begin{gather}
\tilde{\rho}\tilde{T}E\left(\frac{\partial s}{\partial t}+{\bf u}\cdot\nabla s\right)=\frac{E^{2}}{P_{r}}\nabla\cdot(\tilde{\rho}\tilde{T}\nabla s) +\frac{P_r\,c_{o}\,(1-\eta)}{Ra}Q_{\nu}, \label{eq:entropy} 
\end{gather}
where 
\begin{gather}
c_{o} = 2\, \frac{e^{\frac{N_{\rho}}{m}} - 1}{1-\eta^{2}}
\end{gather} 
with $N_{\rho}=\ln(\tilde{\rho}(r_{i})/\tilde{\rho}(r_{o}))$. The viscous heating contribution is given by
\begin{gather}
Q_{\nu}=2\tilde{\rho}\left[ e_{ij} e_{ji} - \frac{1}{3} (\nabla\cdot\mathbf{u})^2 \right].
\end{gather} 
The anelastic approximation also demands that 
\begin{gather}
\nabla\cdot(\tilde{\rho}\mathbf{u})=0.
\end{gather}

The above equations have been non-dimensionalized using the shell thickness $r_o-r_i$ as the length scale,  the inverse rotation rate as the time scale, the entropy contrast $\Delta s$ between top and bottom as the entropy scale, and density and temperature at top boundary as the density and temperature scales. 

Several fundamental control parameters determine the behavior of the above set of equations: the Ekman number $E=\nu/(\Omega d^2)$, the Prandtl number $P_r=\nu/\kappa$, and the Rayleigh number $Ra=g_o d^3\Delta s/(c_p \kappa \nu)$, where $\nu$ is viscosity, $\kappa$ is the thermal diffusivity, $g_o$ is gravity at $r_o$. We assume viscosity and thermal diffusivity to be constant throughout the shell. 

\subsection{Simulation Code}
The hydrodynamic anelastic system of equations are solved using the open-source `MagIC' code ({\tt https://magic-sph.github.io/}) which has been extensively benchmarked against other community codes \cite{jones2011}. It uses a toroidal-poloidal decomposition to maintain strict divergenceless condition where needed, for example, mass flux is given by
\begin{gather}
\tilde{\rho}\mathbf{u}=\nabla\times(\nabla\times W\hat{r}) + \nabla\times X\hat{r}, \nonumber 
\end{gather}
where $W$ and $X$ are scalar potentials. The code is pseudo-spectral in nature and uses spherical harmonic functions horizontally and Chebyshev polynomials radially. The code utilizes the open-source library SHTns \cite{shtns} to perform spherical harmonic transforms. The system of equations is time-advanced using an explicit second-order Adams-Bashforth scheme for Coriolis and non-linear terms and an implicit Crank-Nicolson scheme for the rest of the terms \cite{glatzmaier1984}.

\subsection{Control Parameters}
In this paper, we analyze and report results from one simulation case that acts as a proof-of-concept. The constant non-dimensional control parameters for this case are $E=10^{-5}$, $Pr=0.1$, $Ra=2.3\times10^8$. We span five density scale heights in the simulation, giving a density contrast of about 150 across the shell. The simulation was performed on a grid with 160, 960, 1920 points in $r$, $\theta$, and $\phi$ directions, respectively; the latitude-longitude grid has a maximum spherical harmonic degree of 640. Due to the highly demanding nature of the simulation, we could simulate it for about 0.1 viscous diffusion time (about 1600 rotations), which is similar to  earlier high-resolution studies  \cite{heimpel2005,heimpel2016}. Such a time span is likely not enough to resolve all the available time scales in the system. For instance, the strength and number of zonal jets will evolve on a much longer viscous diffusion time \cite{manfroi1999}. However, the jet meanders and the corresponding vortices (re)form and evolve on the much faster convective turnover time. Furthermore, except for a slow change in the overall zonal flow energy, the kinetic energy also quickly settles to a statistically stationary state. These indicators suggest that the simulation results we discuss below are robust and non-transient phenomenon.

Similar to Heimpel {\it et al.} \cite{heimpel2016} we employ hyperdiffusivity in which the viscosity becomes a function of spherical harmonic degree after a certain cutoff. MagIC code implements it by multiplying the following function to the primary viscous diffusion operator:
\begin{gather}
d(\ell)=1+D\left[\frac{\ell+1-\ell_{hd}}{\ell_{max}+1-\ell_{hd}} \right]^{\beta}
\end{gather}
where $D$ defines the amplitude of the function, $\ell_{max}$ is the maximum spherical harmonic degree utilized in the simulation, $\ell_{hd}$ is the degree after which the hyperdiffusion starts, and $\beta$ defines the rise of the function for degrees higher than $\ell_{hd}$. For our simulation, we use $D=5$, $\beta=5$, $\ell_{hd}=350$.

The boundaries at $r_i$ and $r_o$ are impenetrable and stress-free to the flow. Furthermore, the entropy is assumed constant on each boundary.

\section{Results}

\begin{figure*}
\centering
\includegraphics[width=0.9\linewidth]{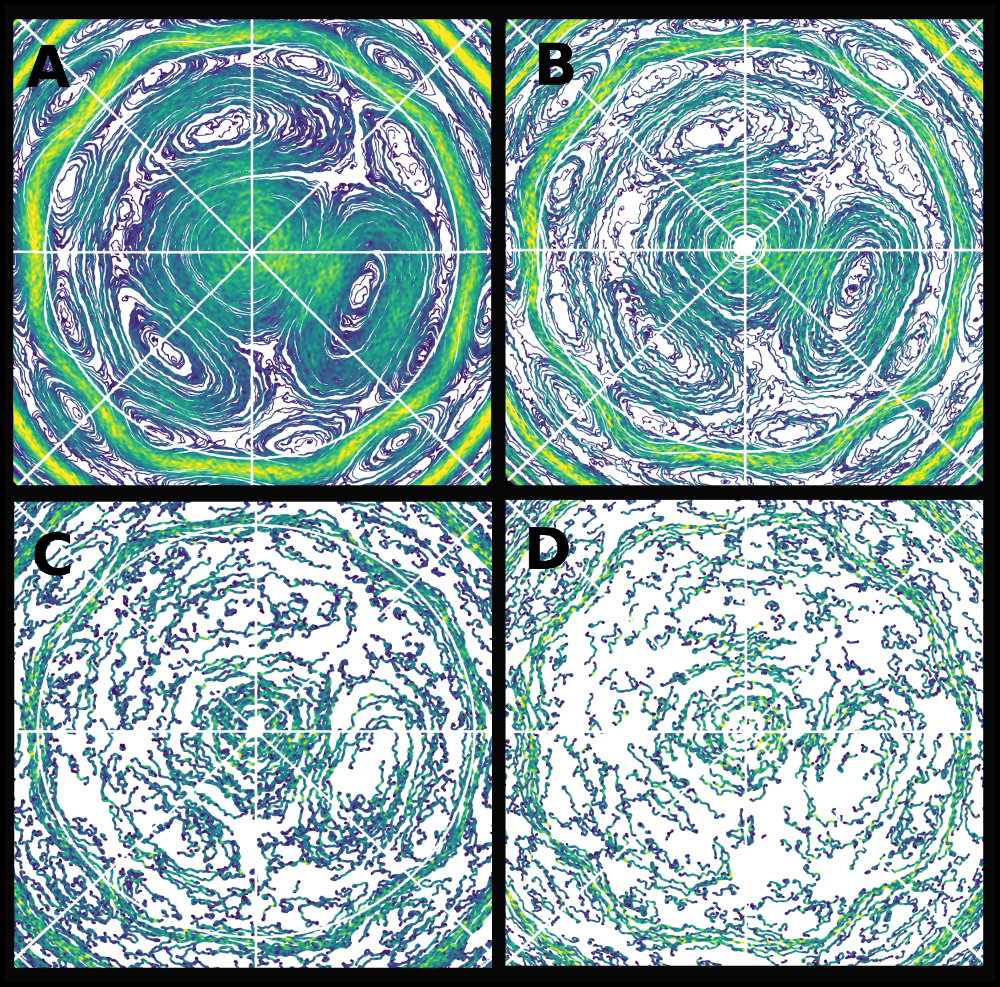}
\caption{Zoomed in stream-line plots of the north polar regions of spherical surfaces at different depths: 0.95$r_o$ in {\bf A}, 0.97$r_o$ in {\bf B}, 0.99$r_o$ in {\bf C}, and 0.993$r_o$ in {\bf D}.  All panels show the simulation case presented in Figure 1 and 2. The figure highlights the gradual emergence of large coherent vortices as well as more coherent jets with depth. Note that the full white circle, representing a constant latitude, is located at 60 degree north of the equator in the {\em local} co-ordinates of each spherical surface. The 60 degree north latitude is omitted in panel {\bf D} for clarity. Due to less coherent flow structures at progressively shallower depths (from panel {\bf A} to {\bf D}), flow stream lines become shorter and less surface filling, exposing more of the hypothetical white surface that is shown for highlighting the coloured stream lines.}
\label{fig3}
\end{figure*}

As the simulation progresses, rotating turbulent convection gradually builds up strong zonal flows. The generated zonal flow profile is shown in Fig.~\ref{fig1}. With the control parameters that we use, the simulation generates a strong prograde jet, up to about 20 degrees from the equator, which is followed by a strong retrograde jet in the vicinity of the tangent cylinder (an imaginary cylindrical surface tangent to the inner boundary and aligned with the rotation axis). Several more alternating jets form at mid and high latitudes. These jets are by far the most energetic component of the flow, carrying more than 10 times the energy contained in the meridional and radial flow components, which is a generic property of simulations with a low enough Ekman number and free slip boundaries \cite{aurnou2007, yadav2016}. The zonal jets in the simulation are qualitatively as well as quantitatively (within a factor of 2) similar to the zonal jets observed on Saturn. Heimpel and Aurnou \cite{heimpel2007} show that the zonal flow velocity and the topographic $\beta$ effect (i.e.~changing axially-vertical fluid column height with latitude) due to the spherical geometry have a large impact on the width and number of zonal jets in such simulations. The fact that the shell thickness and the Rossby number of the zonal flows are similar to the values on Saturn is likely responsible for the good match of the simulation zonal flow profile with the observations.

The zonal jets are largely invariant along the rotation axis (despite a density drop of about 150 across the shell), demonstrating the strong influence of the rotation on the flow at these Rossby numbers. The westward jets north and south of the equator in the simulation are at a somewhat lower latitude than the corresponding jets on Saturn. Noting that these jets usually form in the vicinity of the tangent cylinder in such simulations \cite{heimpel2005,gastine2012,heimpel2016}, we  speculate that  Saturn's atmosphere might extend somewhat deeper than 0.9$R_S$, thereby giving a tangent cylinder at a slightly higher latitude.

To reveal the various dynamical structures present in the simulation, in Fig.~\ref{fig2} we visualize a snapshot of the simulation using flow stream lines on a spherical surface at radius 0.95$r_o$. The figure shows that the system dynamics is much richer than just zonal jets. Along with the jets, there are well defined large-scale vortices at mid and high latitudes. One large cyclonic vortex sits on the pole, accompanied by three anti-cyclonic neighbors. Another set of smaller cyclonic vortices follows to the south of these three anticyclones, followed by a strong eastward jet at about 60 degree north of the equator. In the polar region, (anti)cyclones are arranged such that they roughly define an eastward jet with a triangular pattern. The pattern formed in the 60 degree north jet is one with 9 edges. The other eastward jet visible close to 30 degrees north also contains a polygonal pattern but with a higher wave number and less well defined edges than the jet close to 60 degrees north. The generation of such large scale structures by convection plumes -- whose size is similar to the wiggles in individual stream lines -- shows the presence of an efficient inverse cascade of energy from small to large scales. As mentioned above, the  number and the widths of the zonal jets are likely determined by the zonal flow strength and the topographic $\beta$ effect \cite{heimpel2007}. However, which system parameters control the size and properties of the giant vortices remain unclear and demand a broad control parameter study.

A view of the simulation from a midlatitude vantage point (SI Appendix, Figure S1) reveals that broadly speaking, circular jets dominate at low-latitudes, while large vortices form and appear to induce polygonal shapes in  jets at mid and high latitudes. This is likely driven by the topographical $\beta$ effect, due to the spherical shell geometry, which is stronger near the equator, promoting  strong axisymmetric jets, while, at higher latitudes, the $\beta$-effect decreases and the system approaches a rotating plane layer where the formation of giant vortices is favored (e.g.~see  \cite{rhines1975,cho1996,williams2003,liu2010,heimpel2007,guervilly2014}). 

A remarkable property of the simulation is revealed when we inspect the flow structure as a function of radius in Fig.~\ref{fig3} where stream lines are plotted on spherical surfaces with different depths. As we look at shallower depths, the flow morphology in high latitude regions changes from  smoother stream lines, polar storm, polygonal jet, and coherent large vortices to one with more irregular looking stream lines, a fainter polygonal jet and a central cyclonic storm. We interpret this transition as follows. In density-stratified convection, shallower and lighter fluid must overturn faster to respond to the momentum of fluid parcels coming from deeper thicker layers. This leads to a gradual increase in the mean velocity with increasing radius (e.g.~see \cite{hotta2014, gastine2012}). In the current simulation, the mean velocity increases by a factor of 3 or more in shallower layers (SI Appendix, Figure S2). Therefore, within one simulation, the Rossby number, which depends on the convective time scale (changing with radius) and the rotational time scale (staying constant), changes with depth. This leads to a situation where deeper layers with smaller Rossby number promote more coherent vortices and jets, while shallower layers with larger Rossby number favor more incoherent convection \cite{gastine2012}. This point is elucidated further by the Figure S3 in the SI Appendix which shows regular stream lines in the deep and chaotic ones at shallower depths in a large vortex. Coexistence of both these regimes allow a scenario where the deeper energetic zonal jets manage to extend to the outermost layers, but the large vortices with weaker flow and smaller energy get overpowered by the shallower chaotic convection, and, therefore, loose their identity. The central cyclone, which can be thought of as a tiny zonal jet at the pole, survives since it carries significantly stronger flows than other non-polar vortices. A similar scenario can be imagined for Saturn where the hexagonal shape of the jet is sustained by adjacent six large vortices which are hidden by the more chaotic convection in the shallower layers.

\begin{figure}
\centering
	\includegraphics[width=0.5\textwidth]{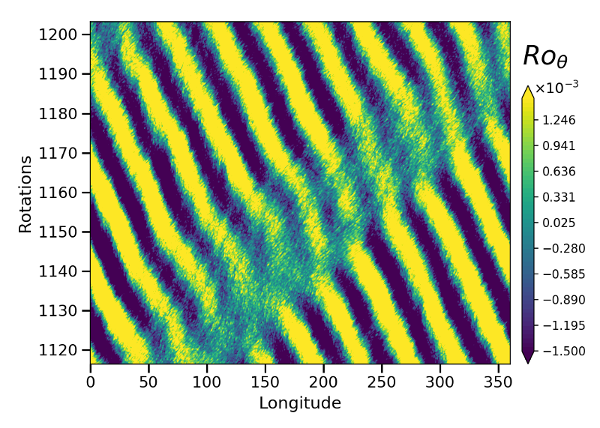}
	\caption{The latitudinal velocity at 55 degrees north of equator as a function of time.} \label{fig4}
\end{figure}

The southern hemisphere of the simulation also exhibits similar flow structures (SI Appendix, Figure S4). However, the precise arrangement is different from the northern region. The polar cyclone is shifted away from the pole and is pinched by two large surrounding anticyclones. Here, too, a polygonal jet exists at around 60 degree south, albeit with polygonal edges only in a limited range of longitudes. Here we note the observation of a similar ephemeral limited-longitude polygonal shape observed by Cassini on Saturn's eastward jet at about 60 degree south \cite{vasavada2006}. The polygonal edges are washed out in the eastward jet close to 30 degrees south. This demonstrates the rich dynamics created by the turbulent fluid interactions. The strength and longitudinal extent of polygonal shapes in the jet is dynamic and evolves (animation showing evolution spanning about 90 rotations is available at \href{https://youtu.be/sfFhAfUT5sI}{https://youtu.be/sfFhAfUT5sI}) due to non-linear interactions. The wavy pattern, as well as the adjacent vortices, drift in the westward direction on the 60 degree north/south eastward jets. Figure \ref{fig4} shows a time evolution of the meridional flow at 55 degree north demonstrating a coherent drift of about -2 degrees per rotation. Saturn's hexagon, on the other hand, is much more stable with -0.06 to 0.01 degree per day \cite{caldwell1993,hueso2019}. The eastward jet close to 30 degree north/south portrays similar behavior; however, this jet has less pronounced and short lived modulations. Since the vortices become less favored at low latitudes, the low latitude jets can indeed be expected to be less influenced by them.

The polygonal modulations in the zonal jets are sensitive to the azimuthal length scale available. When we restrict our simulation domain to only one quarter of the 0 -- 360 degree longitudes, with periodic boundary conditions on the edges, the polygonal patterns disappear and only circular jets remain. Here, the azimuthal length scale of the mid to high latitude zonal jets becomes similar to the length scale of the polygonal modulation. Such conditions do not support the formation of wavy jets in our setup. When we increased the size of the simulated wedge to cover 0 -- 180 degrees longitudes, the polygonal jets appeared again. Furthermore, the polygonal jets are also sensitive to the Rayeligh number of the simulation which sets the mean Rossby number attained in the simulation.  When we decreased the Rayleigh number from $2.3\times 10^8$ (used above) to $1.5 \times 10^8$, the polygonal shapes disappeared and only circular jets remained. When we increased the Rayleigh number to $4.5\times 10^8$ (simulated for about 140 rotations due to computational constraints), the number of polygonal edges decreased from 9 in the case discussed above to 7 (SI Appendix, Figure S5). This trend is similar to those found in earlier laboratory experiment where the wavenumber of the modulation on a zonal jet decreased as the Rossby number of the jet increased \cite{sommeria1989,aguiar2010}. 

\section{Discussion}

Although the model does not capture every aspect of the observations of Saturn, it is, however, the first to produce polygonal zonal jets self-consistently in a deep convection setup. The polygonal shapes form due to mid to high latitude vortices pinching adjacent zonal jets (see Marcus and  Lee \cite{marcus1998} for a similar interpretation). The vortices, however, have much weaker flow (as compared to jets) that gets masked by the more incoherent convection at shallower layers, leaving only polygonal jets as the prevalent flow profile. We find that simulating the entire azimuthal extent of the shell is crucial for modelling meandering jets, which explains why earlier models with wedge simulation geometries did not produce such features. The westward drift of the polygonal shapes was faster (about -2 degree per rotation) than observations in our reported case. However, this drift was significantly lower (about -1.3 degree per rotation) in a simulation at a higher Rayleigh/Rossby number (SI Appendix, Figure S5). Therefore, it is conceivable that a simulation with higher Rossby numbers than what we could achieve will show polygonal jets with much weaker drifts. A more detailed parameter study of the control parameter space should be possible in future with increased computational resources, which will help us narrow down the finer ingredients needed to produce more observations simultaneously in a single model.

\vspace{1cm}

\noindent{\bf Data Availability}: The simulation input file that can be used to reproduce the results is available here: {\tt "https://doi.org/10.6084/m9.figshare.12110982.v1"}. The simulation code used is open access and is available here: {\tt "https://github.com/magic-sph/magic/"}.

\noindent {\bf Acknowledgements}: R.K.Y. thanks Hao Cao for interesting discussions. The work was supported by the NASA Juno project.  The computing resources were  provided by the NASA High-End Computing (HEC) Program through the NASA Advanced Supercomputing (NAS) Division at Ames Research Center and by Research Computing, Faculty of Arts \& Sciences, Harvard University. \\

\bibliographystyle{apsrev4-2}
\bibliography{cited}

\vspace{5cm}

\pagebreak

\makeatletter
\renewcommand{\citenumfont}[1]{S#1}

\renewcommand\thefigure{\arabic{figure}}
\renewcommand{\figurename}{Supplementary Fig.}
\setcounter{figure}{0} 

\begin{figure*}[b]
\includegraphics[scale=0.4]{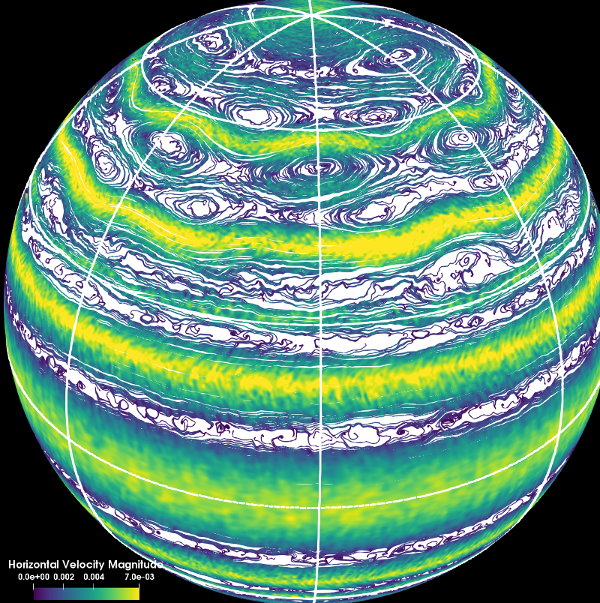}
\caption{\label{EDfig1} Orthographic view from a northern mid-latitude vantage point of the simulation at the same instance shown in Fig.~2 of the main paper.}
\end{figure*}

\begin{figure*}[b]
\includegraphics[scale=0.6]{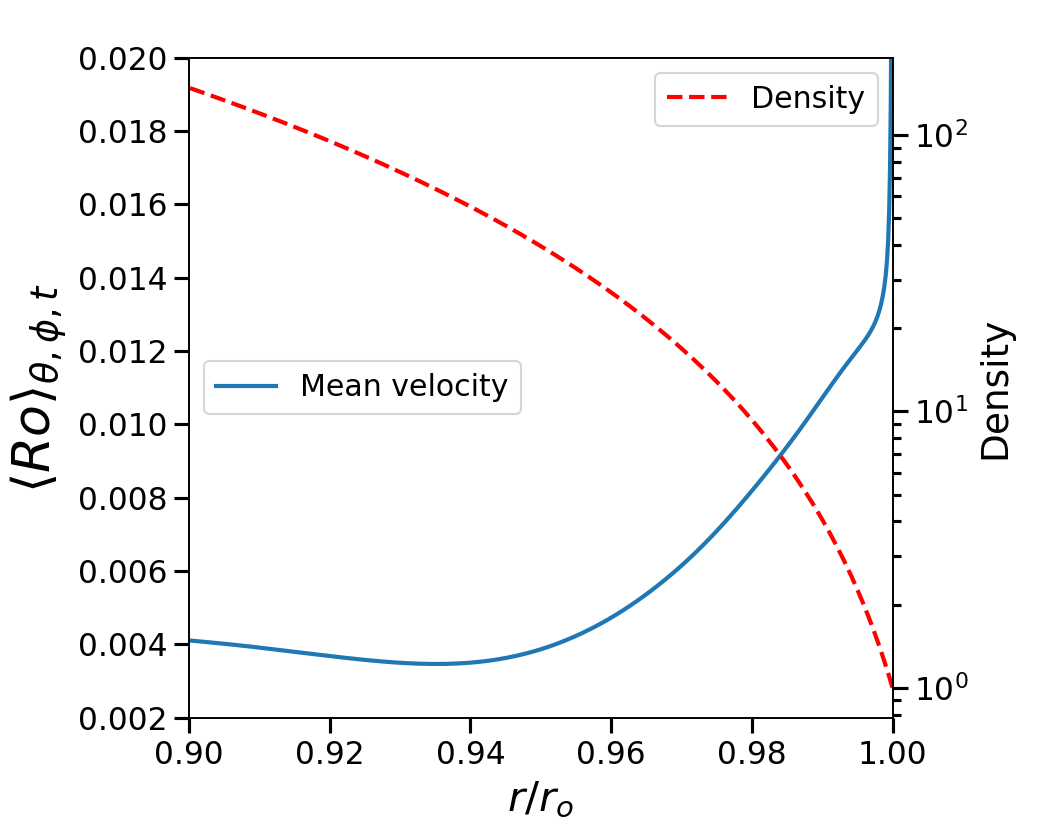}
\caption{\label{EDfig2} Variation of the mean velocity (left axis) expressed in terms of the Rossby number and the normalized fluid density (right axis) as a function of the normalized radius.}
\end{figure*}

\begin{figure*}[b]
\includegraphics[scale=0.9]{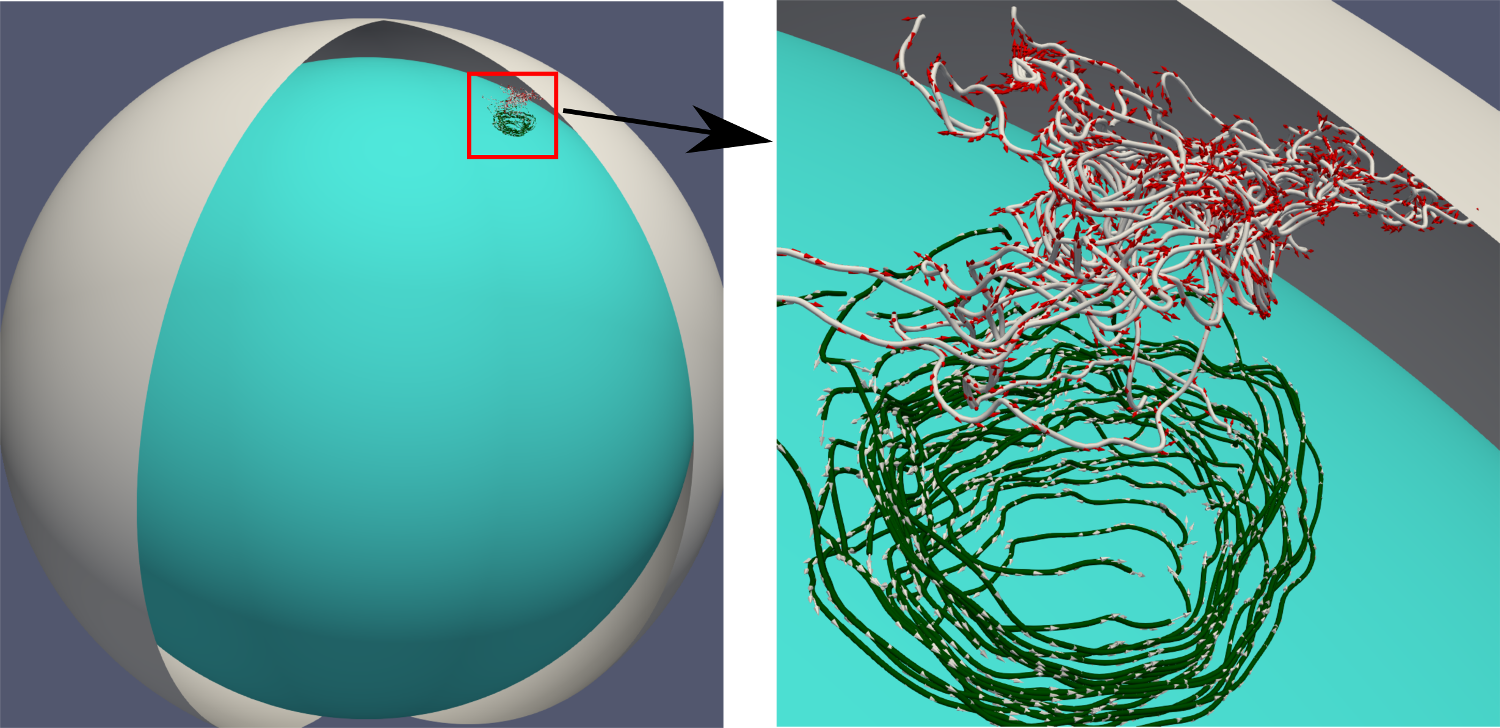}
\caption{\label{EDfig3} Orthographic plot showing gray colored stream lines with red flow vectors for the shallow flow and green colored stream lines with gray flow vectors for deeper flow in a large vortex. The concentric shells show the simulation boundaries.}
\end{figure*}

\begin{figure*}[b]
\includegraphics[scale=1.3]{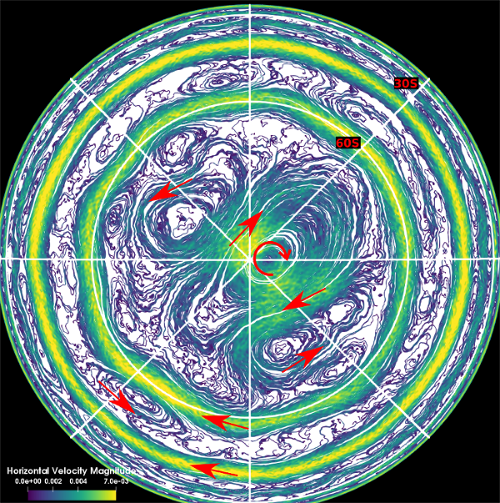}
\caption{\label{EDfig4} Orthographic view from the south pole vantage point of the simulation at the same instance shown in Fig.~2 of the main paper.}
\end{figure*}

\begin{figure*}[b]
\includegraphics[scale=1]{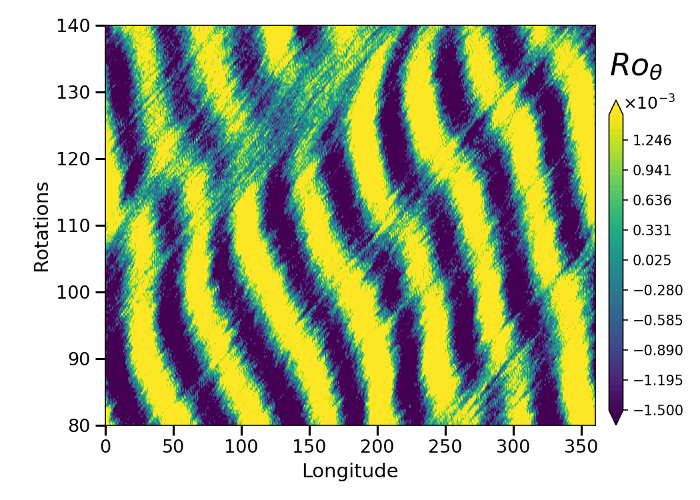}
\caption{\label{EDfig6} The latitudinal velocity at 55 degrees north of equator as a function of time.}
\end{figure*}

\end{document}